\documentclass{article}

\newfont{\msbm}{msbm10}

\def\K{{\cal K}}

\def\A{{\cal A}}

\def\R{\hbox{\msbm R}}

\def\F{{\cal F}}

\newtheorem{prop}{Proposition}
\newtheorem{cor}{Corolary}
\newtheorem{theo}{Theorem}
\newtheorem{deff}{Definition}

\def\bc{\begin{cor}}
\def\ec{\end{cor}}

\def\bt{\begin{theo}}
\def\et{\end{theo}}

\def\bd{\begin{deff}}
\def\ed{\end{deff}}

\def\bp{\begin{prop}}
\def\ep{\end{prop}}

\def\ba{\begin{eqnarray}}
\def\ea{\end{eqnarray}}

\def\be{\begin{equation}}
\def\ee{\end{equation}}

\newfont{\msbms}{msbm6}  

\def\E{{\cal E}}

\def\ab{\mbox{$\bar \A$}}
\def\Si{\mbox{$\Sigma$}}

\usepackage{color}

\begin{document}
\title{Groups of generalized flux transformations \\
in the space of generalized connections}

\author{J. M. Velhinho}

\date{{ Departamento de F\'\i sica, Universidade da Beira 
Interior\\R. Marqu\^es D'\'Avila e Bolama,
6201-001 Covilh\~a, Portugal}\\{jvelhi@ubi.pt}}

\maketitle

\begin{abstract}
\noindent  
We present a group of transformations in the space of generalized connections
that contains the set of transformations generated by the flux variables of
loop quantum gravity. This group is labelled by certain $SU(2)$-valued functions 
on the  bundle of directions in the spatial manifold. A further generalization is obtained by considering functions that depend on germs of analytic curves,
rather than just on directions.
\end{abstract}

\pagestyle{myheadings}



\section{Introduction and motivation}
\label{intro}

The space of generalized connections \ab\ is a distributional 
extension of the space of smooth connections in a manifold~\cite{AL1,AL3,ALMMT,Ba}. 
The space \ab\ features in particular in the canonical approach to quantum gravity 
known as  loop quantum gravity (LQG),\footnote{For reviews of LQG see e.g.~\cite{AL,T}.} where it plays the role 
of a (kinematical) `quantum configuration
space'.

Following early ideas~\cite{AL1,Ba}, a category theory 
description of the space of generalized connections has been  put forward~\cite{Vg,V} 
and explored, aiming at a complete knowledge of the natural algebraic structures present  in \ab. 
As an example of the interest of this type of formulations, we mention the natural emergence of a group of automorphisms
(namely the automorphisms of the groupoid of paths) as a possible
distributional extension of the group of spatial diffeomorphims in LQG.
This extension, proposed in~\cite{V}, was recently investigated
in~\cite{BT},
where the relation with a combinatorial formulation of LQG is also explored.

A still unveiled structure in \ab\ is the group of transformations
generated by the LQG flux variables. Being of the momentum type, the flux 
variables correspond to infinitesimal `translations' in the configuration
space \ab. The one-parameter groups of  transformations generated by each flux
variable are well known~\cite{ST,LOST}, and also both generalizations of these 
transformations~\cite{ST,F2,F3} and categorical formulations~\cite{BT} have been introduced.
However, even including the generalizations, the set of  transformations 
associated with the flux variables does not form a group.
Therefore, in order to accommodate the composition of transformations
generated by flux variables, one needs to consider a class of transformations
in \ab\ more general than those described so far.

Inspired by ideas presented in~\cite{T,ST}, we introduce here
a group
of transformations in \ab\ that includes all the transformations generated 
by the flux variables. Moreover, this group sits inside a larger group
of transformations, of the same type, but depending on a larger amount of information.
All these transformations leave invariant the so-called Ashtekar-Lewandowski measure,
and so they are all unitarily implemented in the LQG (kinematical) Hilbert space.

As in the case of  generalized (local) gauge transformations, the full
group of generalized flux transformations in question can be seen as the action of a 
certain group of functions with values in the gauge group. In the present case, those
functions depend in general on points and on germs of analytic curves in the spatial
manifold.
From a different perspective, this group can also be realized as a subset of \ab,
and in this sense it includes the examples of distributional connections
typically discussed in the literature.

In what follows, we will focus on the trivial bundle case and the analytic set-up. 
To avoid repeating the existing literature (see e.g.~\cite{T} and references therein), only the basic notions from the space
of generalized connections are introduced. 
Also, only the main arguments of proofs are presented, obviating straightforward technical details.  

\section{Generalized flux transformations}
\label{flux}

Let \Si\ be a connected analytic manifold, $G$ a compact, connected 
Lie group,
and $P(\Si,G)$ the trivial principal $G$-bundle over \Si.
($G=SU(2)$ in LQG.)

Let us consider compactly supported, embedded, oriented analytic curves
in \Si, i.e.\ maps $c:[0,1]\to \Si$ whose image is contained in a compact set,
which are analytic in the whole domain, and such that $c(]0,1[)$ is an embedded
submanifold. On the set of these curves one introduces an equivalence relation
which identifies two curves if they differ only by an orientation preserving
analytic reparametrization. Thus, two equivalent curves have  the same image 
and the same orientation. The equivalence classes of such oriented analytic
curves are called {\it edges}. The set of all edges is here denoted by $\E$.

The natural composition of curves leads to a composition of edges, 
$(e_2,e_1)\mapsto e_2 e_1$, if the final point of $e_1$ coincides with the initial
point of $e_2$. The inverse $e^{-1}$ of an edge $e$ is the edge which differs from
$e$ only in the orientation.   

A generalized connection $\bar A$ is map $\bar A:\E\to G$ such that
\be
\label{gc}
\bar A(e^{-1})=\left(\bar A(e)\right)^{-1} \ \ {\rm and}\ \ 
\bar A(e_2 e_1)=\bar A(e_2)\bar A(e_1),
\ee
whenever the composition of edges $e_2 e_1$ is again an edge.\footnote{The composition
of edges does not necessarily produce new edges. In general one obtains
equivalence classes of piecewise analytic curves. The equivalence relation in the set
of piecewise analytic curves includes also modding out by retracings, so that two curves
differing in orientation are inverses of each other. The thus obtained algebraic
structure is a groupoid, and generalized connections are morphisms, 
with values in $G$.}
The set of all such maps $\bar A$ (with an appropriate topology) is the space of generalized connections \ab.

Following~\cite{T,ST}, let us introduce the notion of {\it germ}, for edges. 
Consider the following equivalence
relation in $\E$: two edges are equivalent if they start at the same point and intersect
at an infinite number of points, in which case one of the edges is an analytic
extension of the other. An equivalence class of edges is called a germ.
Thus, a germ at a point $x\in\Si$ is characterized by an infinite number
of Taylor coefficients (in some parametrization) at $x$, from which
the whole family of equivalent edges can be reconstructed.
The set of germs corresponding to a given point $x\in\Si$ is independent
of $x$, and will be denoted by $\K$. The germ of $e$ is denoted by $[e]$.

Since $G$ is a group,  the set $\rm{Map}[\Si\times \K,G]$ of all maps $g:\Si\times \K\to G$ is a  group under pointwise product, i.e. $(gg')(x,[e])=g(x,[e])g'(x,[e])$
defines a group structure. The main object of the present paper is a subgroup
of this group, defined as follows.
\begin{deff}
\begin{itemize}
\item[i)] For a given edge $e$ and a point $x$ on $e$, let $e_x$ denote the subedge of $e$ that starts at $x$, i.e. $e_x$ starts at $x$,
has the same orientation as $e$, and $e_x\cap e=e_x$.
\item[ii)] For a given $g:\Si\times \K\to G$ and an edge $e$, let $S(g,e)$ denote the set of points $x$ along the edge $e$ such that
 $g(x,[e_x])$ is different from the identity of the group $G$.
\item[iii)] The subset $\F\subset \rm{Map}[\Si\times \K,G]$ is defined as 
the set of all elements $g$ such that 
$S(g,e)$ is a finite set for every given\/ edge $e$.
\end{itemize} 
\end{deff}
It is straightforward to see that $\F$ is a subgroup of $\rm{Map}[\Si\times \K,G]$,
since if $S(gg',e)$ is an infinite set for some edge $e$, then either
$S(g,e)$ or  $S(g',e)$ (or both) must be infinite.
\begin{prop} $\F$ is a group with respect to the pointwise product.
\end{prop}

The following are examples of elements of $\F$.  
\begin{itemize}
\item[\bf E1.] $g$ supported on a finite number of points, i.e.
$g$ is different from the identity of $G$ only in a finite number of points
of $\Si$.
On that set, $g$ may  depend arbitrarily on the germs. 
\item[\bf E2.]  $g$ supported
in an analytic line $\ell$, such that $g$ also equals the identity for every germ that defines curves on  the line $\ell$. Appart from that, $g$ may vary from point to point
on $\ell$ and depend on the germ at each point.
\item[\bf E3.]  $g$ supported
in an analytic surface $S$. Like in the {\bf E2} case, $g$ is also required to   
equal the identity for every germ that defines curves on $S$.
\end{itemize}
Examples {\bf E2} and {\bf E3} are well defined  due to the following property:
the number of points $x$ in the intersection
between an analytic edge $e$ and an analytic  surface $S$ (line $\ell$) such that 
$[e_x]$ does not define curves in  $S$ ($\ell$)
is  finite.
The same is true if the surface $S$  (line $\ell$) is of the semianalytic type
considered in~\cite{LOST}. So, $S(g,e)$ is  a finite set in such cases.
(To be precise,  a germ $[e]$ at a point of a surface $S$ defines a curve in $S$
if there is an edge $e$ in the equivalence class $[e]$ such that $e\cap S=e$.)

A subgroup of $\F$ is obtained when we consider functions $g\in\F$ that depend
not on the full information carried by the germ, but only on the tangent direction of the germ, i.e. such that,
at each point, $g(x,[e])=g(x,[e'])$ whenever germs $[e]$ and $[e']$ have the same
tangent direction at the starting point $x$. (More precisely: representative curves $c_e$ and $c_{e'}$, in some parametrization, have the same tangent direction   at the  point $x$.) 
Note that this subgroup
can be defined exactly like $\cal F$, with $\cal K$ being replaced by the set $S^2$ of (oriented) directions
in the tangent space  $T_x\Si$ at a point $x$, as $S^2$ coincides with the 
quotient space of $\K$ by the equivalence relation that identifies germs with the 
same tangent direction.
\begin{deff} The subgroup of $\F$ of those elements $g$ that depend
only on the tangent direction of the germ will be denoted 
by $\cal TF$. 
\end{deff}

Special  elements of $\cal TF$ -- directly related 
to the LQG flux variables -- are determined as follows.
\begin{itemize}
\item[\bf E4.]
Let $S$ be an oriented
analytic 
(or semianalytic) surface and  g a $G$-valued function on $S$. An element $g\in\cal TF$ is obtained by declaring that $g$ is supported on $S$ and, on points $x\in S$:
{\it i)}  $g$  is the identity
for   every germ that defines curves on $S$; {\it ii)} $g(x,[e])={\rm g}(x)$ for germs that define curves pointing upwards (with respect to the orientation of $S$); {\it iii)} $g(x,[e])={\rm g}(x)^{-1}$ for germs that define curves pointing downwards.
\end{itemize}
General elements of $\cal TF$, and among them,  compositions of elements of the type {\bf E4}, will have a more general
dependence on the direction at each point.  

\medskip

Though the following is not the aspect  we are most interested in, it is important
to realize that the full set $\F$ can be  injectively mapped into the space of generalized connections \ab. The image of $\F$ by this map includes all the examples of distributional
connections given in~\cite{AL1,T}.
\begin{prop} There is an injective map $\F\to\ab$, $g\mapsto {\bar A}_g$.
\end{prop}

The images ${\bar A}_g$, which are defined by their actions ${\bar A}_g(e)$ on edges,
are constructed as follows (using the same procedure as for the construction
of graphs adapted to surfaces, or standard graphs~\cite{T}). Given $g\in\F$ and an edge $e$, let us denote the starting
and ending points of $e$ by $x$ and $y$, respectively. The finite sets 
$S(g,e)$ and $S(g,e)\cup\{x,y\}$ are naturally ordered,\footnote{It may happen that
$S(g,e)\cap\{x,y\}\not =\emptyset$, in which case only one copy of $x$ or $y$ is kept
in the set $S(g,e)\cup\{x,y\}$. On the other hand, if $S(g,e)=\emptyset$ then 
${\bar A}_g(e)=\bf 1$.} following the orientation
of $e$. One starts by decomposing $e$ using the set $S(g,e)$, i.e. the edge $e$ is
written as a composition of subdges $e=a_n a_{n-1}\cdots a_1$, where $n$ is some 
integer, $a_k\cap e=a_k$ and the $a_k$'s are edges starting and ending at consecutive points
of the set  $S(g,e)\cup\{x,y\}$, $\forall k=1,\ldots,n$.
Next, if necessary, one further decomposes the edges $a_k$, so that $e$ is
written as a composition of subdges $e=b_m b_{m-1}\cdots b_1$, $m\geq n$, where
each edge $b_k$ either starts or ends (not both) at a point of $S(g,e)$.
In doing so, arbitrary cuts between consecutive points of  $S(g,e)$ are introduced.
Finally, let us replace the edges $b_k$ that end in points of $S(g,e)$ by their
inverses, and rename the set of edges so that the edge $e$ is
written as a composition of subdges
\be
\label{dec}
e=e_m^{\epsilon_m} e_{m-1}^{\epsilon_{m-1}}\cdots e_1^{\epsilon_1},
\ee
where each edge $e_k$ starts at a point of $S(g,e)$ and contains no other point of that
set, and the symbols $\epsilon_k$ take values $\pm 1$.

After a decomposition of the type (\ref{dec}) is performed, ${\bar A}_g(e)$ is simply defined by
\be
\label{gen1}
{\bar A}_g(e)=[{\bar A}_g(e_m)]^{\epsilon_m} [{\bar A}_g(e_{m-1})]^{\epsilon_{m-1}}\cdots [{\bar A}_g(e_1)]^{\epsilon_1},
\ee
\be
\label{gen2}
{\bar A}_g(e_k)=g(x_k,[e_k]),\ \ \ k=1,\ldots,m,
\ee
where   $x_k\in S(g,e)$ is the starting point of the outgoing subedge $e_k$.

The first thing to notice about the above construction is that it is well defined,
although the decomposition (\ref{dec}) is not unique  (due to the arbitrariness
in the choice of the subedges $e_k$). This follows from the fact that ${\bar A}_g(e_k)$ depends only on the germ, and not
on the edge starting at $x_k$.  Secondly, one can easily
check that the maps ${\bar A}_g$ satisfy the properties (\ref{gc}), i.e.
they define elements of \ab.
\medskip

We finally come to the main results of the paper.
\begin{prop} There is a faithful representation, hereby denoted $\Theta$,
of   $\F$ as a group of transformations in \ab.
The  transformations  generated by the LQG flux variables belong to the subgroup 
$\Theta (\cal TF)$ . 
\end{prop}

To construct the representation $\Theta$, it is sufficient to give, for each $g\in \F$,
the images  $\Theta_g(\bar A)$ of every generalized connection, which in turn are defined once $\Theta_g(\bar A)(e)$ is known for every edge. To define $\Theta_g(\bar A)(e)$,
let us  proceed as above, and assume that a decomposition of the type (\ref{dec})
has been performed, corresponding to a given $g\in\F$ and edge $e$. Then
\be
\label{teta1}
\Theta_g(\bar A)(e)=[\Theta_g(\bar A)(e_m)]^{\epsilon_m} [\Theta_g(\bar A)(e_{m-1})]^{\epsilon_{m-1}}\cdots [\Theta_g(\bar A)(e_1)]^{\epsilon_1},
\ee
with 
\be
\label{teta2}
\Theta_g(\bar A)(e_k)={\bar A}(e_k)\, g^{-1}(x_k,[e_k]),\ \ \ k=1,\ldots,m,
\ee
where   $x_k\in S(g,e)$ is the starting point of the edge $e_k$.\footnote{The appearence
of $g^{-1}$ in expression (\ref{teta2}), instead of $g$, is simply related to the 
way in which we write the composition of edges. Also, we are only discussing the nontrivial situation $S(g,e)\not =\emptyset$, otherwise $\Theta_g(\bar A)(e)=\bar A(e)$.} 

As above, one can check that the expressions (\ref{teta1}, \ref{teta2}) are independent
of the particular decomposition (\ref{dec}). 
It is also easy to convince oneself that $\Theta_g(\bar A)$ belongs to \ab,
for every $g$ and $\bar A$ (i.e. that conditions (\ref{gc}) are satisfied), and that
the mapping $g\mapsto \Theta_g$ is injective.

It remains to see that $\Theta$ is a representation. The identity of $\F$ is obviously
mapped to the identity transformation, so let us just consider  the relation 
$\Theta_{gg'}=\Theta_{g}\circ \Theta_{g'}$. For  given $g$ and $g'$, generalized connection $\bar A$ and edge $e$, the relation 
$\Theta_{gg'}(\bar A)(e)=\Theta_{g}\circ \Theta_{g'}(\bar A)(e)$
is clearly satisfied in the two cases $S(g,e)\cap S(g',e)=\emptyset$
and $S(g,e)= S(g',e)$. On the other hand, for each fixed edge $e$, the general case can be reduced
to a combination of the above two cases, and the representation property follows.

Thus, the group $\F$ can be seen as a group of transformations in the space of generalized connections. Let us consider in particular elements $g\in{\cal TF}\subset \F$ of the type {\bf E4}\ above, with the $G$-valued functions ${\rm g}(x)$ being generated by Lie($G$)-valued
functions $f(x)$, i.e. ${\rm g}(x)=\exp(tf(x))$, $t\in\R$. The   transformations $\Theta_g$ corresponding to these elements are precisely the transformations in \ab\ generated by the LQG flux variables (see~equations (17) in~\cite{LOST}). 

Like the transformations generated by the flux variables, the transformations
corresponding to the group $\F$ are continuous with respect to the natural
topology of \ab, and leave the Ashtekar-Lewandowski measure~\cite{AL1} invariant.
The proof is essentially the same as in the case of the 
flux variables~\cite{LOST,F2,OL}.\footnote{For each fixed $g$, every graph in $\Si$ can be split and generators
chosen such that
the transformation $\Theta_g$ is a right translation in the space $G^N$ labelled by the graph.} 
\begin{prop}
The Ashtekar-Lewandowski measure is $\F$-invariant.
\end{prop}

\section{Conclusion}

We have introduced groups of transformations in the space of generalized connections
\ab\ that include all the transformations generated by flux variables.
This constitutes an advance in the characterization and understanding
of the kinematical structure of LQG.

The group $\F$ is an infinite-dimensional and non-abelian
subgroup of a very large product group. The latter is made of copies of the gauge group $G$,
with one copy per each germ of analytic curves and per point in space.
Elements of $\F$ can then be thought of as certain $G$-valued functions
on  the `bundle of germs over the
spatial manifold'. Note that those functions typically equal the identity of $G$ at most points, with non-trivial values occurring  in lower dimensional regions of space. Elements of $\F$ then act on \ab, with the behaviour of generalized connections 
typically  being transformed only over lower dimensional 
regions.\footnote{The action $\Theta$ is most simple in the abelian case
$G=U(1)$. In that situation \ab\ is itself a group, $\F$ is a subgroup of \ab\ and $\Theta$ is the natural action of $\F\subset\ab$ on \ab.}
This means that values $\bar A(e)$ can be affected only for edges intersecting those regions.
A corresponding description of the group $\cal T F\subset\F$ and its action can be made, with the difference that
functions and transformations no longer depend on the full information carried by germs,
but only on the direction.

An important property of the group $\F$ is that its action leaves 
the Ashtekar-Lewandowski measure invariant. It follows
that the standard LQG representation of the holonomy-flux algebra carries an unitary representation of the group of transformations $\Theta(\F)$,
which extends the set of unitary operators corresponding to the 
quantization of the transformations generated by the flux variables.

Even if the exact physical meaning of this unitary representation  of the full group $\Theta(\F)$
is not yet clear, it is our opinion that the group of transformations $\Theta(\F)$
deserves attention and further study, as happened e.g.\ with another extended group
of symmetries of the  Ashtekar-Lewandowski measure, namely the group of automorphisms of the groupoid of paths~\cite{V,BT}. In particular, it is tempting to use  the criteria of unitary implementability
of $\F$, or at least of its subgroup $\cal T \F$,  to explore further the representation theory of the holonomy flux-algebra,
hopefully beyond the beautiful results obtained by Lewandowski, Okolow, Sahlmann,
Thiemann and Fleischhack~\cite{ST,LOST,F2,OL}. In this respect, note that
it already follows from those results (see e.g.~Lemma 6.2 
in~\cite{OL})
that the Ashtekar-Lewandowski measure is the only $\cal T F$-invariant
measure in \ab.\footnote{Measures $\mu$ that are invariant under a transformation $T$
provide a unitary implementation of $T$ in the Hilbert space defined
by $\mu$. The same is true for quasi-invariant measures, i.e.
such that $\mu$ and the push-forward measure $T_*\mu$ are mutually absolutely continuous.}


\section*{Acknowledgements}
\noindent 
I am very greatful to Jos\'e Mour\~ao and Guillermo Mena Marug\'an.
This work was supported in part by 
POCTI/FIS/57547/2004.




\end{document}